\def\e{{\rm e}}
\def\del{\partial}
\def\half{{1\over2}}

\def\vev#1{\langle #1 \rangle}
\def\del{\partial}
\def\half{{1\over2}}

\def\vev#1{\langle #1 \rangle}

\def\del{\partial}
\def\dslash{\del\kern-0.55em\raise 0.14ex\hbox{/}}

\def\rough#1{\raise.3ex\hbox{$#1$\kern-.75em\lower1ex\hbox{$\sim$}}}

\newcommand{\PRD}[3]{{\it Phys. Rev.} {\bf D{#1}}, {#2} (19{#3})}

\newcommand{\PRDM}[3]{{\it Phys. Rev.} {\bf D{#1}}, {#2} (20{#3})}
\newcommand{\PRL}[3]{{\it Phys. Rev. Lett.} {\bf {#1}}, {#2} (19{#3})}

\newcommand{\NPB}[3]{{\it Nucl. Phys.} {\bf B{#1}}, {#2} (19{#3})}
\newcommand{\NPBM}[3]{{\it Nucl. Phys.} {\bf B{#1}}, {#2} (20{#3})}
\newcommand{\PLB}[3]{Phys. Lett. {\bf B{#1}}, {#2} (19{#3})}
\newcommand{\PLBM}[3]{{\it Phys. Lett.} {\bf B{#1}}, {#2} (20{#3})}

\newcommand{\PTP}[3]{{\it Prog. Theor. Phys.} {\bf {#1}}, {#2} (19{#3})}
\newcommand{\PTPM}[3]{{\it Prog. Theor. Phys.} {\bf {#1}}, {#2} (20{#3})}
\newcommand{\ANN}[3]{{\it Ann. Phys. (N.Y.)} {\bf {#1}}, {#2} (19{#3})}

\newcommand{\RMP}[3]{{\it Rev. Mod. Phys.} {\bf{#1}}, {#2} {(19{#3})}}

\documentclass[12pt]{article}
\usepackage{graphicx}
\begin{document}
\baselineskip=18pt
\begin{titlepage}
\begin{flushright}
KOBE-TH-09-05\\
\end{flushright}
\vspace{1cm}
\begin{center}{\Large\bf 
High Temperature Symmetry Nonrestoration and Inverse Symmetry 
Breaking
on Extra Dimensions
}
\end{center}
\vspace{1cm}
\begin{center}
Makoto Sakamoto$^{(a)}$
\footnote{E-mail: dragon@kobe-u.ac.jp} and
Kazunori Takenaga$^{(b)}$
\footnote{E-mail: takenaga@kumamoto-hsu.ac.jp}
\end{center}
\vspace{0.2cm}
\begin{center}
${}^{(a)}$ {\it Department of Physics, Kobe University, 
Rokkodai, Nada, Kobe 657-8501, Japan}
\\[0.2cm]
${}^{(b)}$ {\it Faculty of Health Science, Kumamoto
Health Science University, Izumi-machi, Kumamoto 861-5598, Japan}
\end{center}
\vspace{1cm}
\begin{abstract}
We study $D$-dimensional gauge theory with an extra dimension
of a circle at finite temperature. We mainly focus on the
expectation value of the gauge field for the direction of the 
extra dimension, which is the order parameter of the gauge 
symmetry breaking. We evaluate the effective potential in the 
one-loop approximation at finite temperature. We show that
the vacuum configuration of the theory at finite temperature 
is determined by a $(D-1)$-dimensional gauge theory defined by 
removing the Euclidean time coordinate as well as all of the fermions from
the original $D$-dimensional gauge theory on the circle. 
It is pointed out that gauge symmetry nonrestoration and/or inverse
symmetry breaking can occur at high temperature in a class
of gauge theories on circles and that phase 
transitions (if they occur) are, in general, expected to be the first order. 
\end{abstract}
\end{titlepage}
\newpage
\section{Introduction}
Over the past few decades, a considerable number of studies 
have been made on higher-dimensional field theories where some 
of the spatial coordinates are compactified on certain
topological manifolds. Such theories are found to possess 
unexpectedly rich properties that shed new light and give 
a deep understanding on high energy physics beyond the 
standard model. In fact, it has been shown that new mechanisms
of gauge symmetry breaking \cite{manton, fairlie, sherk,
hosotani, higgsless}, spontaneous supersymmetry 
breaking \cite{susy},
and breaking of translational invariance \cite{translation1,
translation2} can occur, and that various phase structures arise
in field theoretical models
on such 
space-time \cite{higgshosotani1, higgshosotani2}. Furthermore,
new diverse scenarios of solving the hierarchy 
problem have been proposed in \cite{rs, manybrane, 
nagasawasakamoto, sakamototakenaga1}.
\par
Since the origin of gauge symmetry breaking is still an unsolved
problem, it will be worth pursuing alternative possibilities of
gauge symmetry breaking other than the Higgs mechanism. In this
paper, we will focus on the gauge symmetry breaking caused by
expectation values of gauge fields for the directions of extra
dimensions particularly at high temperature. If the gauge
symmetry breaking of $SU(2)\times U(1)$ to $U(1)_{\rm em}$ is 
given by the Higgs mechanism, the gauge symmetry will be restored
at high temperature, since one-loop corrections to the Higgs mass
squared are proportional to $T^2$ with positive coefficients at
high temperature \cite{KL, dolanjackiw, weinberg}. On the other hand, the 
gauge symmetry breaking
via gauge fields for extra dimensions is expected to exhibit
quite different behavior at high temperature. This is because the
UV cutoff dependence is directly related to the high temperature 
corrections \cite{weinberg} and because quantum corrections to 
the zero modes of such gauge fields are expected to be UV finite due to the 
higher-dimensional gauge symmetry (while Higgs mass corrections
are quadratically divergent).
\par
In this paper, we study $D$-dimensional gauge theory with an
extra dimension of a circle at finite temperature, in which 
there are two order parameters of gauge symmetry breaking: One
is the zero mode of the gauge field for the direction of the
Euclidean time, and the other is that for the extra dimension. We
evaluate the effective potential for the zero modes in the
one-loop approximation and clarify the vacuum configuration at
finite temperature. Our results show that at high temperature the
expectation value of the gauge field for the extra dimension in
the $D$-dimensional gauge theory on a circle can be effectively
determined by a $(D-1)$-dimensional theory defined by removing
the Euclidean time coordinate and all of the fermions from the
original $D$-dimensional gauge theory on the circle and that the
gauge symmetry nonrestoration (SNR) and/or the inverse symmetry
breaking (ISB) can occur at high temperature in a class of such
gauge theories. We further point out that phase transitions (if
they occur) are, in general, expected to be first order. It would
be of interest to investigate physical implications of the above
results in cosmology \cite{cosmology}.
\par
This paper is organized as follows: In section $2$, we briefly
explain the zero modes of the gauge fields for the Euclidean
coordinate and the extra dimension, which are the order parameters 
of the gauge symmetry breaking. In section $3$, we evaluate the
one-loop effective potential for the zero modes and write it in
three different ways. In section $4$, we study the expectation
values of the gauge fields in particular at high temperature. In
section $5$, we present five-dimensional $SU(2)$ models with the
gauge symmetry nonrestoration or the inverse symmetry breaking at
high temperature. The final section is devoted to conclusions.
%
%
%
%
%
\section{Dynamical Variables in Finite Temperature Gauge Theories
Compactified on $S^1$}
In this section, we briefly discuss dynamical zero modes, which
are the order parameters of gauge symmetry breaking, 
in $D$-dimensional gauge theory compactified on a circle 
at finite temperature. 
The $D$-dimensional coordinates of the system are decomposed as 
\begin{equation}
x^K=(\tau, x^i,y),\quad (i=1,2,\cdots, D-2),
\label{shiki2-1}
\end{equation}
where $\tau$ denotes the coordinate for the Euclidean time
direction, which is compactified on a circle $S_{\beta}^1$ whose 
circumference is given by the inverse temperature $\beta=T^{-1}$.
The $y$ is the coordinate of the extra dimension, which is
compactified on another circle $S_L^1$ of circumference $L$. The
$x^i (i=1,2,\cdots, D-2)$ are the coordinates on
$(D-2)$-dimensional flat Euclidean space $R^{D-2}$. According to
the decomposition (\ref{shiki2-1}) of the coordinates, the gauge
fields $A_K$ are also decomposed as 
\begin{equation}
A_K=(A_{\tau}, A_i, A_y),\quad (i=1, 2,\cdots, D-2).
\label{shiki2-2}
\end{equation}
Since the system has two distinct circles of $S_{\beta}^1$ and
$S_L^1$, the zero modes of $A_{\tau}$ and $A_y$, {\it i.e.} the 
expectation values $\vev{A_{\tau}}$ and $\vev{A_y}$ become
dynamical variables, which cannot be gauged away, because the
circles are multiply-connected \cite{hosotani}.
\par
A physical consequence of nontrivial vacuum expectation values  
$\vev{A_{\tau}}$ and $\vev{A_y}$ is to make the gauge bosons 
massive through the couplings
\begin{equation}
g^2\left({\rm tr}[\vev{A_{\tau}},~\vev{A_i}]^2
+{\rm tr}[\vev{A_y},~ \vev{A_i}]^2
\right).
\label{shiki2-3}
\end{equation}
The appearance of massive gauge bosons is a signal for the gauge
symmetry breaking, so that the gauge symmetry breaking patterns
are determined by $\vev{A_{\tau}}$ and $\vev{A_y}$.
\par
It should be noted that the tree-level potential from the
coupling ${\rm tr}F_{\tau y}^2$ arises as 
\begin{equation}
V_{tree}=g^2{\rm tr}[\vev{A_{\tau}},~\vev{A_y}]^2.
\label{shiki2-4}
\end{equation}
In terms of the dimensionless order parameters
\begin{equation}
\vev{a_{\tau}}\equiv g\vev{A_{\tau}}{\beta\over 2\pi},
\quad
\vev{a_y}\equiv g\vev{A_y}{L\over 2\pi},
\label{shiki2-5}
\end{equation}
which are more suitable for our discussions, the tree potential 
(\ref{shiki2-4}) can be written as 
\begin{equation}
V_{tree}={(2\pi)^4\over g^2\beta^2L^2}
{\rm tr}[\vev{a_{\tau}},~ \vev{a_y}]^2.
\label{shiki2-6}
\end{equation}
We observe that in the weak coupling limit, the term
(\ref{shiki2-6}) dominates the potential, so that it is natural to
expect that the vacuum configuration lies along the flat
direction 
\begin{equation}
[\vev{a_{\tau}},~\vev{a_y}]=0.
\label{shiki2-7}
\end{equation}
In the following, we assume the relation (\ref{shiki2-7}).
%
%
%
%
%
\section{One-Loop Effective Potential}
In this section, we evaluate the one-loop effective potential 
for the dynamical variables $\vev{a_{\tau}}$ and $\vev{a_y}$ in
the $D$-dimensional finite temperature gauge theory with an extra
dimension of the circle $S_L^1$. We show that the effective
potential can be written into three different expressions: 
One is suitable for examining the behavior of high temperature or
a large extra dimension. Another is suitable for the opposite limit
of low temperature or a small extra dimension. 
The third one is useful for numerical computations. 
\par
Before driving the effective potential for $\vev{a_{\tau}}$ and
$\vev{a_y}$, we will first give the effective potential for 
$\vev{a_{\tau}}$ in $D$-dimensional gauge theory at finite
temperature but without compactification, and then that 
for $\vev{a_{\tau}}$ at zero temperature with the circle
compactification. 
The forms of the effective potentials without
compactification or at zero temperature turn out to be helpful in
understanding the high or low temperature behavior of the finite
temperature gauge theory with the extra dimension.
%
%
%
\subsection{Effective Potential at Finite Temperature}
Let us first consider $D$-dimensional finite temperature gauge
theory without compactification. In this system, the dynamical
order parameter is given by $\vev{a_{\tau}}$. The standard
prescription \cite{gpy} to evaluate effective potentials for
$\vev{a_{\tau}}$ leads to \cite{sakamototakenaga2}
\footnote{The one-loop effective potential for $\vev{A_{\tau}}$
has been derived in the $SU(N)$ gauge 
theory at finite temperature \cite{weiss, KL2001, FP}. The
one-loop effective potential in the scenario of gauge-Higgs
unification at finite temperature has been studied in \cite{marutake}.}
\begin{equation}
V^D(\vev{a_{\tau}}, \beta, M)_{\cal R}=
{\cal N}(-1)^{f+1}{2\over (2\pi)^{{D\over 2}}}\sum_{l=1}^{\infty}
\left({M\over \beta l}\right)^{D\over 2}
K_{{D\over 2}}(M\beta l)
{\rm tr}_{({\cal R})}
\biggl[\cos\biggl(2\pi l(\vev{a_{\tau}}+\eta)\biggr)\biggr]. 
\label{shiki3-1}
\end{equation}
This is a general form of the effective potentials associated
with a one-loop diagram in which a particle with bulk mass $M$
propagates. The superscript $D$ stands for the number of the
total dimensions. The $\cal N$ is the number of (on-shell)
degrees of freedom for the particle. For example, ${\cal N}=1, 2,
2^{[{D\over 2}]}$, and $2^{[{D\over 2}]-1}$ for a real scalar, a
complex scalar, a Dirac spinor, and a Weyl spinor, respectively. The
factor $(-1)^{f+1}$ comes from the loop of the diagram and
$(-1)^{f+1}=-1~(+1)$ for bosons (fermions). The $\eta$ is defined
by $\eta=0~(\half)$ for bosons (fermions). The origin of this
phase is the quantum statistics for bosons and fermions: Any 
bosonic (fermionic) fields have to obey the periodic
(antiperiodic) boundary conditions with respect to the Euclidean
time coordinate $\tau$, {\it i.e.}
\begin{equation}
\phi(\tau+\beta)=\left\{
\begin{array}{ll}
+\phi(\tau) &\mbox{for~bosons},\\[0.3cm]
-\phi(\tau) &\mbox{for~fermions.}
\end{array}
\right.
\label{shiki3-2}
\end{equation}
When the particle $\phi$ which propagates the loop belongs to the
representation $\cal R$ of the gauge group,
$\vev{a_{\tau}}=\vev{a_{\tau}^c}T_{\cal R}^c$, where $T_{\cal R}^c$
denotes a generator of the gauge group in the representation
$\cal R$. Thus, the trace on (\ref{shiki3-1}) should be taken
over the gauge indices with respect to the representation $\cal
R$. The $K_{\nu}(z)$ is the modified Bessel function defined by
\begin{equation}
\int_0^{\infty}dt~t^{-\nu-1}\e^{-At-{B\over t}}=
2\left({A\over B}\right)^{\nu\over 2}K_{\nu}(2\sqrt{AB}).
\label{shiki3-3}
\end{equation} 
It should be noted that the mode $l$ in (\ref{shiki3-1})
corresponds to the winding number around the circle $S_{\beta}^1$
but not the Matsubara frequency mode. The winding modes are
introduced from the Matsubara modes through the Poisson summation
formula. Actually, we will see the inverse process later.
\par
For the massless case, the potential (\ref{shiki3-1}) reduces to
\begin{equation}
V^D(\vev{a_{\tau}}, \beta, M=0)_{\cal R}=
{\cal N}(-1)^{f+1}{\Gamma({D\over 2})\over \pi^{D\over 2}\beta^D}
\sum_{l=1}^{\infty}{1\over l^D}
{\rm tr}_{({\cal R})}
\biggl[\cos\biggl(2\pi l(\vev{a_{\tau}}+\eta)\biggr)\biggr],
\label{shiki3-4}
\end{equation}
where we have used the formula
\begin{equation}
\lim_{z\rightarrow 0}z^{\nu}K_{\nu}(z)=2^{\nu-1}\Gamma(\nu).
\label{shiki3-5}
\end{equation}
\par
For the fundamental representation of $SU(N)$, the effective
potential (\ref{shiki3-1}) can be written in the familiar
expression 
\begin{equation}
V^D(\vev{a_{\tau}},\beta;M)_{fund}=
{\cal N}(-1)^{f+1}{2\over (2\pi)^{D\over 2}}
\sum_{l=1}^{\infty}
\left({M\over \beta l}\right)^{D\over 2}K_{D\over 2}(M\beta l)
\sum_{i=1}^N\cos\biggl(2\pi l(\varphi_i+\eta)\biggr),
\label{shiki3-6}
\end{equation}
where the expectation value $\vev{a_{\tau}}$ is diagonalized by
an appropriate (constant) gauge transformation as
$
\vev{a_{\tau}}={\rm diag.}(\varphi_1,\varphi_2,\cdots, \varphi_N)
$
with $\sum_{i=1}^N\varphi_i=0$. For the adjoint 
representation of $SU(N)$, Eq. (\ref{shiki3-1}) is written as
\begin{equation}
V^D(\vev{a_{\tau}},\beta;M)_{adj}=
{\cal N}(-1)^{f+1}{2\over (2\pi)^{D\over 2}}
\sum_{l=1}^{\infty}
\left({M\over \beta l}\right)^{D\over 2}K_{D\over 2}(M\beta l)
\sum_{i,  j=1}^N\cos\biggl(2\pi l(\varphi_i-\varphi_j+\eta)\biggr).
\label{shiki3-7}
\end{equation}
%
%
%
\subsection{Effective Potential on a Circle}
In the following, we consider $D$-dimensional gauge theory
on the circle $S_L^1$ at zero temperature. The one-loop effective
potential is given by \cite{hosotani, pomarol, takenaga} 
\begin{equation}
V^D(\vev{a_y}, L, \alpha ;M)=
{\cal N}(-1)^{f+1}{2\over (2\pi)^{D\over 2}}
\sum_{n=1}^{\infty}\left({M\over Ln}\right)^{D\over 2}
K_{D\over 2}(MLn)
{\rm tr}_{({\cal R})}
\biggl[\cos\biggl(2\pi n(\vev{a_{y}}+\alpha)\biggr)\biggr].
\label{shiki3-8}
\end{equation}
In this system, the dynamical order parameter is the expectation
value of the gauge field $A_y$ ( or $a_y$) for the extra
dimension and the inverse temperature $\beta$ in (\ref{shiki3-1})
should be replaced by the circumference $L$ of the circle $S_L^1$.
A difference between (\ref{shiki3-1}) and (\ref{shiki3-8}) arises
in the phases of $\eta$ and $\alpha$. It should be emphasized
that the boundary conditions for the Euclidean time direction are
uniquely determined by the quantum statistics, while those for
the spatial extra dimension are not, {\it a priori}, known and
can be, in general, twisted as 
\begin{equation}
\phi(y+L)=\e^{i2\pi \alpha}\phi(y).
\label{shiki3-9}
\end{equation}
\par
Note that the mode $n$ in (\ref{shiki3-8}) corresponds to the
winding number around the circle $S_L^1$ for the spatial extra
dimension but not the Kaluza-Klein mode. The winding modes are 
derived from the Kaluza-Klein modes through the Poisson summation
formula. Actually, we will see the inverse process later. The
expressions for $M=0$ and for the fundamental and adjoint
representations of $SU(N)$ will be obtained similarly as before.
%
%
%
\subsection{Effective Potential for $\vev{a_{\tau}}$ and $\vev{a_y}$}
Let us finally examine $D$-dimensional gauge theory with an extra 
dimension of the circle at finite temperature. Since the system
possesses two circles of $S_{\beta}^1$ and $S_L^1$, both of the
expectation values $a_{\tau}$ and $a_y$ become
dynamical. According to the standard prescription, a general
form of one-loop effective potentials is found to 
be of the form \cite{sakamototakenaga2}
\begin{eqnarray}
V^D(\vev{a_{\tau}},\beta;\vev{a_y}, L,\alpha;M)_{\cal R}
&=&{\cal N}(-1)^{f+1}{2\over (2\pi)^{D\over 2}}{\rm tr}_{({\cal R})}
\biggl[
\nonumber\\
&&\sum_{l=1}^{\infty}\left({M\over \beta l}\right)^{D\over 2}
K_{D\over 2}(M\beta l)\cos\biggl(2\pi l(\vev{a_{\tau}}+\eta)\biggr)
\nonumber\\
&+&\sum_{n=1}^{\infty}\left({M\over Ln}\right)^{D\over 2}
K_{D\over 2}(MLn)\cos\biggl(2\pi n(\vev{a_y}+\alpha)\biggr)
\nonumber\\
&+&2\sum_{l=1}^{\infty}\sum_{n=1}^{\infty}
\left({M\over {\sqrt{(\beta l)^2+(Ln)^2}}}\right)^{D\over 2}
K_{D\over 2}\biggl(M\sqrt{(\beta l)^2+(Ln)^2}\biggr)
\nonumber\\
&&\times 
\cos\biggl(2\pi l(\vev{a_{\tau}}+\eta)\biggr)
\cos\biggl(2\pi n(\vev{a_y}+\alpha)\biggr)
\biggr],
\label{shiki3-10}
\end{eqnarray}
where we have used the 
relation (\ref{shiki2-7}). Since the expression (\ref{shiki3-10}) is 
suitable for investigating neither the high temperature behavior nor the
low temperature one, we rewrite it into two other different
expressions. This is the main purpose of this subsection.
\par
To this end, we first combine the second and the third terms in 
(\ref{shiki3-10}) together:
\begin{eqnarray}
\mbox{2nd}&+&\mbox{3rd~terms~in}~(\ref{shiki3-10})
\nonumber\\
&=&{\cal N}(-1)^{f+1}{2\over(2\pi)^{D\over 2}}{\rm tr}_{({\cal R})}
\biggl[
\nonumber\\
&&
\sum_{l=-\infty}^{\infty}\sum_{n=1}^{\infty}
\Biggl({M\over\sqrt{(\beta l)^2+(Ln)^2}}\Biggr)^{D\over 2}
K_{D\over 2}\biggl(M\sqrt{(\beta l)^2+(Ln)^2}\biggr)
\nonumber\\
&&\times
\e^{i2\pi l(\vev{a_{\tau}}+\eta)}
\cos\biggl(2\pi n(\vev{a_{y}}+\alpha)\biggr)
\biggr]
\nonumber\\
&=&
{\cal N}(-1)^{f+1}{1\over(4\pi)^{D\over 2}}
{\rm tr}_{({\cal R})}\biggl[
\nonumber\\
&&\sum_{l=-\infty}^{\infty}\sum_{n=1}^{\infty}
\int_0^{\infty}dt~t^{-{D\over 2}-1}~\e^{-{1\over 4t}
((\beta l)^2+(Ln)^2)-M^2t}
~\e^{i2\pi l(\vev{a_{\tau}}+\eta)}
\nonumber\\
&&\times
\cos\biggl(2\pi n(\vev{a_{y}}+\alpha)\biggr)
\biggr]\nonumber\\
&=&{\cal N}(-1)^{f+1}
{1\over(4\pi)^{D-1\over 2}}{\rm tr}_{({\cal R})}
\biggl[
\nonumber\\
&&{1\over \beta}
\sum_{{\tilde l}=-\infty}^{\infty}\sum_{n=1}^{\infty}
\int_0^{\infty}dt~t^{-{D-1\over 2}-1}
~\e^{-{(Ln)^2\over 4t}-t[M^2+({2\pi\over \beta})^2
({\tilde l}+\vev{a_{\tau}}+\eta)^2]}\nonumber\\
&&\times \cos\biggl(2\pi n(\vev{a_{y}}+\alpha)\biggr)
\biggr]\nonumber\\
&=&
{\cal N}(-1)^{f+1}{2\over (2\pi)^{{D-1\over 2}}}
{\rm tr}_{({\cal R})}\biggl[
\nonumber\\
&&{1\over \beta}\sum_{{\tilde l}=-\infty}^{\infty}\sum_{n=1}^{\infty}
\left({M_{\tilde l}\over Ln}\right)^{D-1\over 2}
K_{D-1\over 2}(M_{\tilde l}Ln)
\cos\biggl(2\pi n(\vev{a_{y}}+\alpha)\biggr)\biggr],
\label{shiki3-11}
\end{eqnarray}
where 
\begin{equation}
M_{\tilde l}\equiv \sqrt{M^2+\left({2\pi\over \beta}\right)^2({\tilde l}
+\vev{a_{\tau}}+\eta)^2}.
\label{shiki3-12}
\end{equation}
Note that $M_{\tilde l}$ is identical to the mass of the
Matsubara mode ${\tilde l}$ at finite temperature. 
In the second and the third
equalities in (\ref{shiki3-11}), we have used the formula 
(\ref{shiki3-3}) and the Poisson summation formula 
\begin{equation}
\sum_{l=-\infty}^{\infty}\e^{-{(\beta l)^2\over 4t}+
i2\pi l(\vev{a_{\tau}}+\eta)}=
{\sqrt{4\pi t}\over \beta}\sum_{{\tilde l}=-\infty}^{\infty}
\e^{-t({2\pi\over \beta})^2({\tilde l}+\vev{a_{\tau}}+\eta)^2},
\label{shiki3-13}
\end{equation}
respectively. In the last equality in (\ref{shiki3-11}), the
formula (\ref{shiki3-3}) has been used again.
\par
Inserting (\ref{shiki3-11}) into (\ref{shiki3-10}), we 
have
\begin{eqnarray}
&&V^D(\vev{a_{\tau}}, \beta; \vev{a_y}, L, \alpha;M)_{\cal R}
\nonumber\\
&=&{\cal N}(-1)^{f+1}{2\over (2\pi)^{D\over 2}}
{\rm tr}_{({\cal R})}\biggl[
\sum_{l=1}^{\infty}\left({M\over\beta l}\right)^{D\over 2}
K_{D\over 2}(M\beta l)\cos\biggl(2\pi l(\vev{a_{\tau}}+\eta)\biggr)
\biggr]\nonumber\\
&&+{1\over\beta}\sum_{{\tilde l}=-\infty}^{\infty}
{\cal N}(-1)^{f+1}{2\over (2\pi)^{D-1\over 2}}
{\rm tr}_{({\cal R})}
\biggl[
\nonumber\\
&&
\sum_{n=1}^{\infty}
\left({M_{\tilde l}\over Ln}\right)^{D-1\over 2}
K_{D-1\over 2}(M_{\tilde l}Ln)
\cos\biggl(2\pi n(\vev{a_y}+\alpha)\biggr)
\biggr].
\label{shiki3-14}
\end{eqnarray}
In terms of the effective potential (\ref{shiki3-1}) and 
(\ref{shiki3-8}), we find that the effective potential
(\ref{shiki3-14}) can be represented as 
\begin{equation}
V^D(\vev{a_{\tau}},\beta;\vev{a_y}, L, \alpha ;M)_{\cal R}
=V^D(\vev{a_{\tau}}, \beta;M)_{\cal R}+
{1\over \beta}\sum_{{\tilde l}=-\infty}^{\infty}
V^{D-1}(\vev{a_{y}},L,\alpha; M_{\tilde l})_{\cal R}.
\label{shiki3-15}
\end{equation}
This expression turns out to be suitable for studying the
behavior of high temperature or a large extra dimension, as we will
see in the next section. It should be emphasized that the formula
(\ref{shiki3-15}) has a clear physical interpretation: The
effective potential of the $D$-dimensional gauge theory on
$S_L^1$ at finite temperature $T$ is given by the sum of the
$D$-dimensional effective potential at finite temperature without
compactification and the (one-dimensional lower)
$(D-1)$-dimensional effective potentials on $S_L^1$ at $T=0$ for
Matsubara modes with masses $M_{\tilde l}$ (times $1/\beta$).
\par
Let us next rewrite (\ref{shiki3-10}) into another expression. To
this end, we combine the first and the third terms in
(\ref{shiki3-10}) together:
\begin{eqnarray}
\mbox{1st}&+&\mbox{3rd~terms~in}~(\ref{shiki3-10})
\nonumber\\
&=&{\cal N}(-1)^{f+1}{2\over(2\pi)^{D\over 2}}{\rm tr}_{({\cal R})}
\biggl[
\nonumber\\
&&
\sum_{l=1}^{\infty}\sum_{n=-\infty}^{\infty}
\Biggl({M\over\sqrt{(\beta l)^2+(Ln)^2}}\Biggr)^{D\over 2}
K_{D\over 2}\biggl(M\sqrt{(\beta l)^2+(Ln)^2}\biggr)
\nonumber\\
&&\times
\e^{i2\pi n(\vev{a_y}+\alpha)}
\cos\biggl(2\pi l(\vev{a_{\tau}}+\eta)\biggr)
\biggr]
\nonumber\\
&=&
{\cal N}(-1)^{f+1}{1\over(4\pi)^{D\over 2}}
{\rm tr}_{({\cal R})}\biggl[
\nonumber\\
&&\sum_{l=1}^{\infty}\sum_{n=-\infty}^{\infty}
\int_0^{\infty}dt~t^{-{D\over 2}-1}~\e^{-{1\over 4t}
((\beta l)^2+(Ln)^2)-M^2t}
~\e^{i2\pi n(\vev{a_y}+\alpha)}
\nonumber\\
&&\times
\cos\biggl(2\pi l(\vev{a_{\tau}}+\eta)\biggr)
\biggr]\nonumber\\
&=&{\cal N}(-1)^{f+1}
{1\over(4\pi)^{D-1\over 2}}{\rm tr}_{({\cal R})}
\biggl[
\nonumber\\
&&{1\over L}
\sum_{{\tilde n}=-\infty}^{\infty}\sum_{l=1}^{\infty}
\int_0^{\infty}dt~t^{-{D-1\over 2}-1}
~\e^{-{(\beta l)^2\over 4t}-t[M^2+({2\pi\over L})^2
({\tilde n}+\vev{a_y}+\alpha)^2]}\nonumber\\
&&\times \cos\biggl(2\pi l(\vev{a_{\tau}}+\eta)\biggr)
\biggr]\nonumber\\
&=&
{\cal N}(-1)^{f+1}{2\over (2\pi)^{{D-1\over 2}}}
{\rm tr}_{({\cal R})}\biggl[
\nonumber\\
&&{1\over L}\sum_{{\tilde n}=-\infty}^{\infty}\sum_{l=1}^{\infty}
\left({M_{\tilde n}\over \beta l}\right)^{D-1\over 2}
K_{D-1\over 2}(M_{\tilde n}\beta l)
\cos\biggl(2\pi l(\vev{a_{\tau}}+\eta)\biggr)\biggr],
\label{shiki3-16}
\end{eqnarray}
where 
\begin{equation}
M_{\tilde n}\equiv \sqrt{M^2+\left({2\pi\over L}\right)^2({\tilde n}
+\vev{a_y}+\alpha)^2}.
\label{shiki3-17}
\end{equation}
We should note that $M_{\tilde n}$ is nothing but the mass of the
Kaluza-Klein mode $\tilde n$ associated with the circle
compactification $S_L^1$. In the second and the third equalities in
(\ref{shiki3-16}), we have used the formula (\ref{shiki3-3}) and
the Poisson summation formula
\begin{equation}
\sum_{n=-\infty}^{\infty}\e^{-{(Ln)^2\over 4t}
+i2\pi n(\vev{a_y}+\alpha)}
={\sqrt{4\pi t}\over L}\sum_{{\tilde n}=-\infty}^{\infty}
\e^{-t{({2\pi\over L})^2({\tilde n}+\vev{a_y}+\alpha)^2}},
\label{shiki3-18}
\end{equation}
respectively. In the last equality in (\ref{shiki3-16}), we
have again used the formula (\ref{shiki3-3}).
\par
Inserting (\ref{shiki3-16}) into (\ref{shiki3-10}), we find 
\begin{equation}
V^D(\vev{a_{\tau}},\beta; \vev{a_y},L,\alpha;M)_{\cal R}
=V^D(\vev{a_y}, L, \alpha; M)_{\cal R}+
{1\over L}\sum_{{\tilde n}=-\infty}^{\infty}
V^{D-1}(\vev{a_{\tau}},\beta; M_{\tilde n})_{\cal R}.
\label{shiki3-19}
\end{equation}
This expression turns out to be suitable for studying the
behavior of low temperature or a small extra dimension, as we will
see in the next section. It follows from the relation
(\ref{shiki3-19}) that the effective potential of the
$D$-dimensional gauge theory on $S_L^1$ at finite temperature $T$
is found to be equivalent to the  sum of the $D$-dimensional 
effective potential on $S_L^1$ at zero temperature and the (one-dimension
lower) $(D-1)$-dimensional effective potentials at finite
temperature $T$ (without compactification) for the Kaluza-Klein
modes of masses $M_{\tilde n}$ (times $1/L$).
%
%
%
%
%
\section{Symmetry Nonrestoration and Inverse Symmetry Breaking at
High Temperature}
We have succeeded in deriving the three different expressions for
the one-loop effective potential of the $D$-dimensional gauge
theory on $S_L^1$ at finite temperature $T$. Using those results,
we clarify the vacuum structure for the order parameters
$\vev{a_{\tau}}$ and $\vev{a_y}$ at high temperature as well as
low temperature, and show that the symmetry nonrestoration and/or
inverse symmetry breaking can occur at high temperature in gauge
theories with an extra dimension of a circle.
\par
Let us first investigate the vacuum configuration at high
temperature, {\it i.e.} $LT\gg 1$. It turns out that the second
expression (\ref{shiki3-15}) (or (\ref{shiki3-14})) is
particularly suitable for that purpose. In the high temperature
limit of $LT\rightarrow \infty$, the first term in
(\ref{shiki3-15}) (or (\ref{shiki3-14})) is dominant because the
first term is proportional to $T^D$, while the second one is
proportional to $T$. The first term is the one-loop effective
potential of the $D$-dimensional uncompactified gauge theory at
finite temperature and turns out to determine the expectation
value $\vev{a_{\tau}}$ (but not $\vev{a_y}$). 
In \cite{sakamototakenaga2}, the
vacuum configuration of $\vev{a_{\tau}}$ has been extensively
studied in finite temperature gauge theories. It has been shown
that $\vev{a_{\tau}}$ cannot acquire nontrivial vacuum expectation
values in finite temperature $SU(N)$ gauge theories that consist
of an arbitrary number of matter fields belonging to the fundamental
and the adjoint 
representations \footnote{There is an exception. If the models 
includes only the matter belonging to
the adjoint representation of $SU(N)$, $\vev{a_{\tau}}$ can take
one of the values in the center of $SU(N)$, {\it i.e.}
$\e^{i2\pi\vev{a_{\tau}}}=\e^{i{2\pi\over N}k}
{\bf 1}_{N\times N}~(k=0, 1, \cdots, N-1)$ due to the ${\bf Z}_N$
symmetry \cite{sakamototakenaga2}}, as well as finite 
temperature $SU(2)$ gauge theories without any restrictions on 
the matter contents. Hence, the
analyses strongly suggest that $\vev{a_{\tau}}$ is trivial, {\it
i.e.} $\vev{a_{\tau}}=0$. Although we take $\vev{a_{\tau}}$ to be
zero in the following discussions, we will arrive at similar
conclusions even if $\vev{a_{\tau}}\neq 0$. 
\par
Putting $\vev{a_{\tau}}=0$, we find that the expectation value
$\vev{a_y}$ can be determined by the second term in
(\ref{shiki3-15}). Since the modified Bessel function
$K_{\nu}(z)$ has an asymptotic form $\sqrt{{\pi\over 2z}}\e^{-z}$
as $z\rightarrow \infty$, it exponentially decreases to zero 
as $z\rightarrow \infty$. This immediately implies that all of the
fermions as well as nonzero Matsubara modes $({\tilde l}\neq 0)$
do not contribute to the second term in (\ref{shiki3-15})
because $M_{\tilde l}L\gg 1$ for $LT\gg 1$ 
with $\eta = {1\over 2}$ or ${\tilde l}\neq 0$. 
Hence, we can write the total effective
potential for $\vev{a_y}$ symbolically as 
\begin{equation}
V^D(\vev{a_y};\beta, L)_{total}
\stackrel{LT\gg 1}{\simeq}{1\over \beta}
V^{D-1}(\vev{a_y};L)\bigg|_{{without~fermions}}.
\label{shiki4-1}
\end{equation}
Since the overall factor $\beta^{-1}$ is irrelevant to the
determination of $\vev{a_y}$, we arrive at an important
conclusion that at high temperature $LT\gg 1$ the expectation
value $\vev{a_y}$ is determined by a $(D-1)$-dimensional theory
defined by removing the Euclidean time coordinate as well as all
of the fermions \footnote{Ghost fields should not be excluded because
they obey the periodic boundary conditions for the Euclidean time
direction \cite{hatakugo}.} from the original $D$-dimensional 
finite temperature gauge theory on the circle.
\par
We should make a comment on the above dimensional reduction at
high temperature. It is known that at high temperature the
thermal properties of a $(3+1)$-dimensional field theory are given
by an effective three-dimensional field 
theory \cite{ginsparg, bn}. However, our
present situation is quite different from that 
given in \cite{ginsparg, bn}: All of the fermions are completely 
decoupled at high temperature, but bosons
are not. The effective potential (\ref{shiki4-1}) is essentially
independent of temperature $T$ at $LT\gg 1$ since the overall
factor $\beta^{-1}=T$ does not affect the minimization procedure
of the potential and 
since $V^{D-1}(\vev{a_y};L)|_{\rm without~fermions}$ is completely 
independent of $T$. Thus, the temperature
dependence of the effective potential for $\vev{a_y}$ disappears
at high temperature up to an overall factor $\beta^{-1}$. This is
in sharp contrast to effective potentials in ordinary finite
temperature field theories, where masses or couplings for Higgs
fields are, in general, temperature-dependent at high
temperature.
\par
Let us next discuss the vacuum configuration of the theory at low
temperature, {\it i.e.} $LT\ll 1$. In this case, the third
expression (\ref{shiki3-19}) is found to be suitable. At low
temperature $LT \ll 1$, the first term in (\ref{shiki3-19})
becomes dominant because the first term is proportional to
$L^{-D}$, while the second one is proportional to $L^{-1}$. Thus,
the expectation value $\vev{a_y}$ is determined by the effective
potential of the original $D$-dimensional gauge theory
compactified on $S_L^1$ at $T=0$, as expected naively.  
\par
The vacuum expectation value $\vev{a_{\tau}}$ can then be
determined by the second term in (\ref{shiki3-19}) with a
``background'' $\vev{a_y}$. Since our previous analyses 
\cite{sakamototakenaga2} strongly
suggest that $\vev{a_{\tau}}$ acquires no nontrivial expectation
values, we thus arrive at the conclusion that at low temperature
(or a small extra dimension) $LT\ll 1$ the vacuum configuration of
the theory is determined by the original $D$-dimensional gauge
theory on $S_L^1$ at zero temperature.
\par
We have shown that the expectation value $\vev{a_y}$ of the
$D$-dimensional gauge theory on $S_L^1$ at high temperature
$LT\gg 1$ is given by that of a $(D-1)$-dimensional theory 
defined by removing the Euclidean time coordinate and all of the
fermions from the original $D$-dimensional gauge theory, while at
low temperature $LT\ll 1$, $\vev{a_y}$ is given by that of the
original $D$-dimensional gauge theory on $S_L^1$ at zero temperature.
A crucial observation is that the fermion contribution disappears
in determining $\vev{a_{y}}$ at high temperature, so that we
expect that the vacuum configuration at high temperature can be
different from that at low temperature. It is known that
$\vev{a_y}$ can acquire nontrivial expectation values even
without fermions by appropriately choosing representations of the
gauge group and twist parameters $\alpha$'s for scalars. This
fact immediately suggests that the symmetry
nonrestoration or inverse symmetry breaking can occur at high
temperature in a class of gauge theories with extra dimensions
compactified on circles \footnote{Models with SNR or ISB have
been previously reported in multi-$\lambda\phi^4$ models
\cite{weinberg} and little Higgs models \cite{littleHiggs}.}. This 
is indeed the case. 
\par
In the next section, we will present $5d~SU(2)$ gauge models 
compactified on a circle with the gauge symmetry nonrestoration 
or the inverse symmetry breaking at high temperature, as a 
demonstration. We would like to point out that phase transitions 
in such a class of gauge theories are expected to be, in 
general, first order. This observation comes from the fact 
that effective potentials for $\vev{a_y}$ include higher powers 
of $\vev{a_y}$ and also odd powers of it. In the next 
section, we will see that the model with ISB causes the first 
order phase transition at a critical temperature. 
%
%
%
%
%
\section{$5d~SU(2)$ Models with SNR and ISB}
In this section, we construct $5d$ finite temperature $SU(2)$
gauge models on a circle with the gauge symmetry nonrestoration
and the inverse symmetry breaking at high temperature. 
We numerically study the behavior of the effective 
potential with respect to the temperature for certain matter 
content. We assume all of the matter fields are massless, so 
that (\ref{shiki3-10}), which is suitable for the numerical 
analyses, becomes
\begin{eqnarray}
&&V^{D=5}(\vev{a_{\tau}}, \vev{a_y}, t, \alpha; M=0)_{\cal R}
\nonumber\\
&=&{\cal N}(-1)^{f+1}{\Gamma({5\over 2})\over \pi^{5\over 2}L^5}
{\rm tr}_{({\cal R})}\biggl[
\sum_{n=1}^{\infty}{1\over n^5}
\cos\biggl(2\pi n(\vev{a_{y}}+\alpha)\biggr)
+
t^5~\sum_{l=1}^{\infty}{1\over l^5}
\cos\biggl(2\pi l(\vev{a_{\tau}}+\eta)\biggr)
\nonumber\\
&&+
2~t^5~\sum_{l=1}^{\infty}\sum_{n=1}^{\infty}
{1\over{[l^2+(tn)^2]^{5\over 2}}}
\cos\biggl(2\pi l(\vev{a_{\tau}}+\eta)\biggr)
\cos\biggl(2\pi n(\vev{a_{y}}+\alpha)\biggr)
\biggr],
\label{shiki5-1}
\end{eqnarray}
where we have defined the dimensionless parameter
$t\equiv LT$ and have used (\ref{shiki3-5}). The 
two order parameters of gauge symmetry breaking are 
\begin{equation}
\vev{a_y}={\rm diag.}(\theta, -\theta),\quad
\vev{a_{\tau}}={\rm diag.}(\varphi, -\varphi).
\label{shiki5-2}
\end{equation} 
for the gauge group $SU(2)$. Once we fix the matter content, the
effective potential is given by summing each contribution of the 
effective potential (\ref{shiki5-1}) from the gauge and the
matter fields. The vacuum configuration $\theta$ is determined 
by minimizing the total effective potential for fixed values 
of $LT$ and $\varphi=0$, and we then find the unbroken gauge symmetry 
of the model.
\par
It has been known that the matter content and the boundary conditions
of fields for the $S_L^1$ direction are crucial 
for the determination of $\vev{a_y}$ \cite{models}. In order to break 
the original gauge symmetry through the nontrivial value of $\vev{a_y}$, adjoint 
fermions and/or twisted scalars are necessary in the concerned
model. On the other hand, the scalars satisfying the periodic boundary
condition do not break the gauge symmetry. Since we have already
understood the dominant contributions to the effective potential at
high (low) temperature, one can appropriately choose the matter
content in such a way to realize the SNR/ISB. 
\par
Let us first present a model with the gauge symmetry 
nonrestoration. We consider the matter content given by
\begin{equation}
\left(N_{fd}^s,~\alpha\right)=\left(2,~\half\right),
\left(N_{adj}^s,~\alpha\right)=\left(6,~\half\right),
\left(N_{fd}^f,~\alpha\right)=\left(1,~0\right),
\left(N_{adj}^f,~\alpha\right)=\left(2,~0\right),
\label{shiki5-3}
\end{equation}
where $N_{adj}^{s(f)}$ and $N_{fd}^{s(f)}$ stand for the number of 
the adjoint scalars (fermions) and for the one of the fundamental scalars
(fermions), respectively. The $\alpha$ denotes the twisted boundary
condition defined by (\ref{shiki3-9}). We consider the scalars
satisfying the antiperiodic boundary condition in the model.
\par
At zero temperature, $LT=0$, the vacuum configuration of the 
model is numerically found to be  
\begin{equation}
\theta\simeq 0.277459.
\label{shiki5-4}
\end{equation} 
The $SU(2)$ gauge symmetry is broken down to $U(1)$. 
Then we turn on and increase the temperature. We 
depict the behavior of the vacuum expectation 
value $\theta$ with respect to $LT$ in Fig.$1$ by 
numerical calculations. 
\begin{figure}[ht]
\begin{center}
\includegraphics[width=9cm]{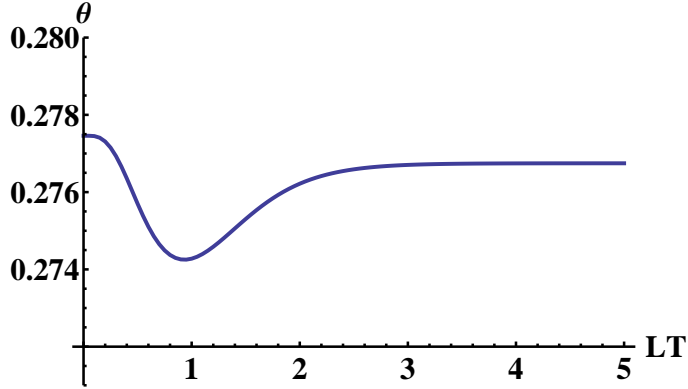}
\caption{The behavior of $\theta$ with respect to
$LT$ in the model (\ref{shiki5-3}). The asymptotic value 
of $\theta$ for $LT\gg 1$ is about $0.276746$, which is 
consistent with the one obtained from the
four-dimensional gauge theory explained in the text.}
\end{center}
\end{figure}
We do not find the vacuum configuration that respects the
$SU(2)$ gauge symmetry for any values of $LT$.
This means that, even if we go to the high temperature, the gauge
symmetry is not restored, unlike the usual Higgs mechanism.
Hence, the model realizes the symmetry nonrestoration. At 
high temperature, $LT\gg 1$, we find that the vacuum 
expectation value approaches
\begin{equation}
\theta\simeq 0.276746.
\label{shiki5-5}
\end{equation}
We also observe that the model does not have a phase transition.
\par
As discussed in the previous section, at high 
temperature, $LT\gg 1$, $\theta$ is determined by
a four-dimensional $(R^3\times S_L^1)$ gauge theory 
including the bosonic degrees of freedom alone. In the high 
temperature limit, the present model (\ref{shiki5-3}) consists 
of the gauge, adjoint and fundamental scalar fields.
The vacuum expectation value $\theta$ of the four-dimensional gauge 
theory is numerically found to be $\theta\simeq 0.276746$, which is the 
same as (\ref{shiki5-5}) obtained by the original 
five-dimensional gauge theory in the high temperature limit. 
\par
Let us next present a model with the inverse symmetry
breaking. The matter content in this case is given by
\begin{equation}
\left(N_{fd}^s,~\alpha\right)=\left(1,~\half\right),
\left(N_{adj}^s,~\alpha\right)=\left(5,~\half\right),
\left(N_{fd}^f,~\alpha\right)=\left(1,~0\right),
\left(N_{adj}^f,~\alpha\right)=\left(0,~0\right).
\label{shiki5-6}
\end{equation}
The notations are the same as before.
Let us note again that we consider the scalars
satisfying the antiperiodic boundary condition. 
\par
At zero temperature, $LT=0$, the vacuum configuration of the
model is found to be 
\begin{equation}
\theta=0.5,
\label{shiki5-7}
\end{equation}
for which the $SU(2)$ gauge symmetry is unbroken. Let us note
that the massless gauge boson exists in the spectrum even 
for the nontrivial value (\ref{shiki5-7}). When we turn on and 
increase the temperature, the vacuum expectation value $\theta$
changes as depicted in Fig.$2$
\begin{figure}[ht]
\begin{center}
\includegraphics[width=9cm]{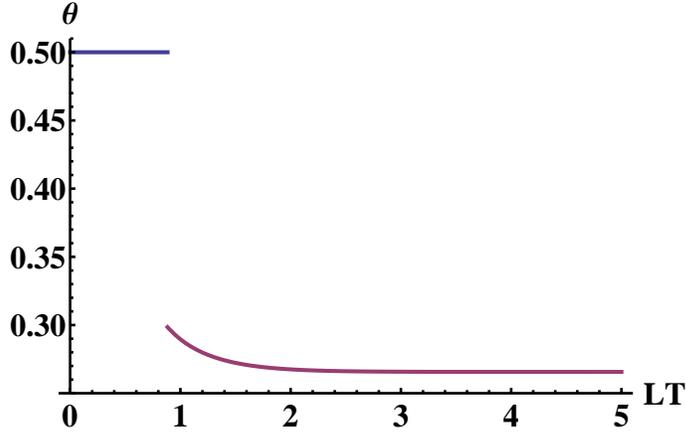}
\caption{The behavior of $\theta$ with respect to $LT$
in the model (\ref{shiki5-6}). The critical 
temperature, at which the vacuum expectation value 
jumps, is $LT_c\simeq 0.8836$. The asymptotic value of
$\theta$ is about $0.265649$, which is consistent with
the value obtained by the four-dimensional gauge theory explained
in the text.}
\end{center}
\end{figure}
We observe that the value of $\theta$ jumps at the critical
temperature $LT_c\simeq 0.8836$, and the degenerate vacuum
configurations---one is $\theta=0.5$, and the other is
$\theta\simeq 0.298456$---appear. The latter
configuration breaks the $SU(2)$ gauge symmetry down to $U(1)$.
The asymptotic value for $LT\gg 1$ is 
\begin{equation}
\theta\simeq 0.265649.
\label{shiki5-8}
\end{equation} 
The $SU(2)$ gauge symmetry is not restored. Hence, the model 
realizes the inverse symmetry breaking.
We also depict the behavior of the effective potential at around
the critical temperature in Fig. $3$. The phase transition is
clearly first order, which is also understood from the
discontinuity of the vacuum expectation value in Fig. $2$. 
\begin{figure}[ht]
\begin{center}
\includegraphics[width=9cm]{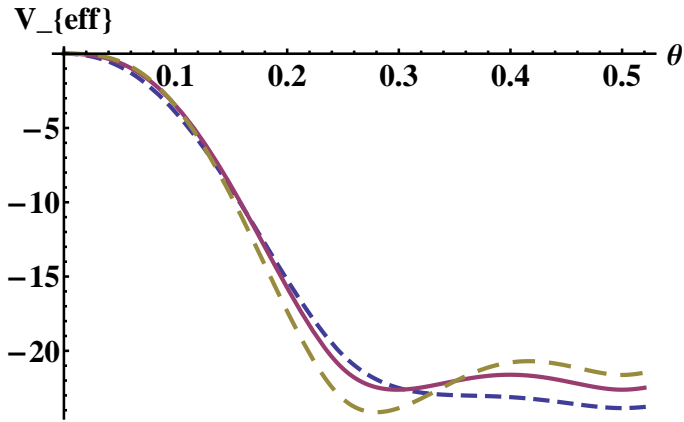}
\caption{The behavior of the effective potential with respect to
$LT$ in the model (\ref{shiki5-6}). The solid (dashed, dotted)
line corresponds to $LT=LT_c (1.2, 0.5)$, respectively. The 
critical temperature $LT_c$ is about $0.8836$. The potential is 
symmetric with respect to $\theta=0.5$, as understood
from (\ref{shiki5-1}) and (\ref{shiki5-2}).}
\end{center}
\end{figure}
\par
At high temperature, $LT\gg 1$, the $\theta$ is determined by
a four-dimensional $(R^3\times S_L^1)$ gauge theory, where 
all of the fermions are decoupled and the Euclidean time 
coordinate shrinks. The $\theta$ in the four-dimensional 
gauge theory is numerically found to 
be $\theta\simeq 0.265649$. This is the 
same as (\ref{shiki5-8}) obtained by the original 
five-dimensional gauge theory at the high temperature limit. 
\par
We have presented the models with the symmetry 
nonrestoration and the inverse symmetry breaking
at high temperature. We note that at high temperature
the model is described by the four-dimensional $(R^3\times S_L^1)$
gauge theory including only bosonic degrees of freedom. The 
nontrivial boundary condition $\alpha$ for the scalar fields is 
crucial for the two interesting phenomena at high temperature. 
\par
Let us finally present a model which has both the usual 
symmetry restoration of the $SU(2)$ gauge symmetry 
and the inverse symmetry breaking in the intermediate
range of $LT$. The matter content is given by
\begin{equation}
\left(N_{fd}^s,~\alpha\right)=\left(2,~\half\right),
\left(N_{adj}^s,~\alpha\right)=\left(6,~\half\right),
\left(N_{fd}^f,~\alpha\right)=\left(1,~0\right),
\left(N_{adj}^f,~\alpha\right)=\left(0,~0\right).
\label{shiki5-9}
\end{equation}
The vacuum configuration at zero temperature is numerically found
to be 
\begin{equation}
\theta\simeq 0.359407,
\label{shiki5-10}
\end{equation}
for which the $SU(2)$ gauge symmetry is broken to $U(1)$. If we 
increase the temperature, the behavior of the vacuum expectation value
$\theta$ with respect to $LT$ is given in Fig. $4$.
\begin{figure}[ht]
\begin{center}
\includegraphics[width=9cm]{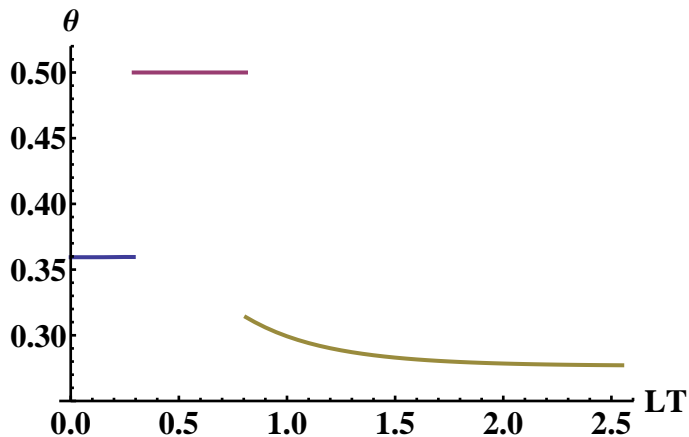}
\caption{The behavior of $\theta$ respect to
$LT$ in the model (\ref{shiki5-9}). The first (second) jump of
the vacuum expectation value 
occurs at $LT\simeq 0.2914 (0.8104)$. The asymptotic value for
$LT\gg 1$ is about $0.276746$, which is consistent with the one
obtained by the four-dimensional gauge theory explained in the text.}
\end{center}
\end{figure}
The first order phase transition occurs 
at $LT_{c1}\simeq 0.2914$, and the $SU(2)$ gauge symmetry is 
restored like the usual Higgs mechanism. The second phase
transition, which is again first order, occurs at 
$LT_{c2}\simeq 0.8104$. The $SU(2)$ gauge symmetry is again
broken to $U(1)$ at high temperature. The 
asymptotic value 
for $LT\gg 1$ is $\theta \simeq 0.276747$. This model
belongs to the 
class of the inverse symmetry breaking at high temperature, but
the model has the usual symmetry restoration of the $SU(2)$ gauge
symmetry in the intermediate range of the temperature. This 
is similar to the phase transition discussed in \cite{littleHiggs}.
%
%
%
%
%
\section{Conclusions}
We have investigated the $D$-dimensional gauge theory with an
extra dimension of a circle, in particular at high and low
temperature. The theory possesses two order parameters of gauge
symmetry breaking: One is the zero mode of the gauge field for
the Euclidean time direction and the other is that for the
direction of the extra dimension. We have evaluated  the
effective potential for the order parameters in the one-loop 
approximation and succeeded in expressing it in three different
forms: One is suitable for high temperature/large extra
dimension, another is suitable for low temperature/a small extra
dimension, and the third one is useful for numerical
computations. Using those expressions and mainly focusing on the
expectation value $\vev{a_y}$ of the extra component of the gauge
field, we have found that at high temperature the effective
potential for $\vev{a_y}$ in the $D$-dimensional gauge theory on
the circle reduces to that in a $(D-1)$-dimensional theory
defined by removing the Euclidean time coordinate and all of the
fermions from the original theory, while at low temperature it is
determined by the original $D$-dimensional gauge theory on the
circle at zero temperature. This result leads to the conclusion
that the gauge symmetry nonrestoration and/or the inverse
symmetry breaking at high temperature can occur in a class of
gauge theories on a circle with appropriate matter contents. 
\par
As a demonstration, we have presented $5d~SU(2)$ gauge models 
on a circle that cause the gauge symmetry nonrestoration or 
the inverse symmetry breaking to indeed occur at high 
temperature. Those properties should be contrasted to the gauge 
symmetry breaking induced by the Higgs mechanism. In this 
case, one-loop radiative corrections to the Higgs mass 
squared will be proportional to
$T^2$ with positive coefficients at high temperature, so that 
the gauge symmetry is expected to be restored at high
temperature \cite{KL, dolanjackiw, weinberg}.
We have further discussed that phase transitions in such a class of
gauge theories are, in general, first order and have shown
that it is actually the case for the models we have studied.  
\par
Considering the higher rank gauge group such as $SU(3)$ is one of 
the simplest extensions of our studies. There are more 
symmetry breaking patterns through $\vev{a_y}$ for that case than the
case of $SU(2)$. We expect a rich variety of vacuum structure at 
finite temperature \cite{sakatakenew}. 
\par
We can also easily extend the expressions of (\ref{shiki3-15}) and 
(\ref{shiki3-19}) to gauge theories with extra dimensions of a
higher-dimensional tori $T^p=S_{L_1}^1\times S_{L_2}^1\times
\cdots \times S_{L_p}^1$. Then the one-loop effective potential
can be expressed in the following two ways:
\begin{eqnarray}
V^D
(\vev{a_{\tau}}, \beta ; \vev{a_{y_i}}, L_i, 
\alpha_i ;M)_{\cal R}
&=&V^D(\vev{a_{\tau}}, \beta; M)_{\cal R}
+{1\over\beta}\sum_{{\tilde l}=-\infty}^{\infty}
V^{D-1}(\vev{a_{y_i}}, L_i, \alpha_i; M_{\tilde l})_{\cal R}
\nonumber\\
&=&V^D(\vev{a_{y_i}}, L_i, \alpha_i; M)_{\cal R}
\nonumber\\
&&+{1\over L_1\cdots L_p}
\sum_{{\tilde n}_1,\cdots,{\tilde n}_p=-\infty}^{\infty}
V^{D-p}
(\vev{a_{\tau}}, \beta; M_{{\tilde n}_1
\cdots{\tilde n}_p})_{\cal R},
\label{shiki6-1}
\end{eqnarray}
where
\begin{equation}
M_{{\tilde n}_1\cdots{\tilde n}_p}\equiv
\sqrt{M^2+\sum_{i=1}^{p}\left({2\pi\over L_i}\right)^2
({\tilde n}_i+\vev{a_{y_i}}+\alpha_i)^2}.
\label{shiki6-2}
\end{equation}
It immediately follows that at high temperature the vacuum
configuration for $\vev{a_{y_i}} (i=1,\cdots, p)$ can be
determined by the $(D-1)$-dimensional (zero temperature) gauge
theory on the tori $T^p$ without fermions, while at low temperature it can
be determined by the original $D$-dimensional zero temperature
gauge theory on $T^p$, so that the gauge symmetry nonrestoration
and/or the inverse symmetry breaking at high temperature can
occur in a class of gauge theories compactified on $T^p$.  
\par
We would like to make some comments on high and low temperature
approximations and also higher order effects. We have shown that the
one-loop effective potential (\ref{shiki3-10}) can be written into the
two other different forms (\ref{shiki3-15}) and (\ref{shiki3-19}).
The expression (\ref{shiki3-15}) ((\ref{shiki3-19})) is useful for the
analysis at high (low) temperature $LT\gg 1$ ($LT\ll 1$) because
contributions of the nonzero Matsubara (Kaluza-Klein) modes are 
exponentially suppressed. Those modes, however, become important at
$LT\approx 1$, and high (low) temperature approximations will break down.
Then the analyses of the one-loop effective potential 
at $LT\approx 1$ may be performed by numerical computations with the
original expression (\ref{shiki3-10}). The above observation suggests
that a phase transition will occur at a critical temperature $T_c$ of
order $1/L$ if the high temperature vacuum configuration is different from
the low temperature one. This has been confirmed for the models studied 
in the section $5$.  
\par
Our considerations are restricted to the one-loop
approximation. Higher order effects could alter one-loop results. One
such effect will be temperature-dependent mass 
corrections. At two-loop order, the squared mass $M_{\tilde l}^2$ 
of the Matsubara mode ${\tilde l}$ could acquire a mass correction 
$\Delta M^2\sim g^2T^3$ \footnote{Note that the five-dimensional gauge
coupling constant $g^2$ has mass dimension $-1$, so that $g^2T^3$
has mass dimension $2$.}. Then $M_{\tilde l}$ in
(\ref{shiki3-14}) is expected to be replaced by 
$\sqrt{M_{\tilde l}^2+\Delta M^2}$. If $\sqrt{\Delta M^2}L\gg 1$, {\it i.e.}
$LT \gg 1/(g_4^2)^{1/3}$, where $g_4^2\equiv g^2/L$ is a $4d$
dimensionless coupling constant, contributions of all of the bosonic
modes would be exponentially suppressed like fermionic
ones. Therefore, for such a high temperature region, higher order
effects should be considered properly.
\par
Our results suggest that if the gauge symmetry breaking was
caused not by Higgs fields but by gauge fields of extra dimensions,
the gauge symmetry could not be restored at the early Universe or the
gauge symmetry could be restored with the first order phase
transition. It would be of great importance to investigate
cosmological implications in such a scenario.
%
%
%
%
\begin{center}
{\bf Acknowledgement}
\end{center}
This work is supported in part by a Grant-in-Aid for Scientific Research
(No. 18540275 (M.S.) and No. 21540285 (K.T.)) from the Japanese 
Ministry of Education, Science, Sports and Culture. 
The authors thank Y. Hosotani, T. Inagaki, Y. Kikukawa, T. Onogi, 
H. So, H. Sonoda and H. Yoneyama for valuable discussions.
\vspace*{1cm}
%
%
%
%
%

%
%
%
%
%

\begin{thebibliography}{99}
\bibitem{manton}
M. S. Manton, \NPB{158}{141}{79}.
\bibitem{fairlie}
D. B. Fairlie, \PLB{82}{97}{79}.
\bibitem{sherk}
J. Scherk and J. Schwarz, \PLB{82}{60}{79}; \NPB{153}{61}{79}.
\bibitem{hosotani}
Y. Hosotani, \PLB{126}{309}{83}, \ANN{190}{233}{89}.
\bibitem{higgsless}
C. Csaki, C. Grojean, H. Murayama, L. Pilo and J. Terning,
\PRDM{69}{055006}{04}.
\bibitem{susy}
M. Sakamoto, M. Tachibana and K. Takenaga, \PLB{458}{231}{99}; 
\PTPM{104}{633}{00}.
\bibitem{translation1}
M. Sakamoto, M. Tachibana and K. Takenaga, \PLB{457}{33}{99}.
\bibitem{translation2}
S. Matsumoto, M. Sakamoto and S. Tanimura, \PLBM{518}{163}{01};
M. Sakamoto and S. Tanimura, \PRDM{65}{065004}{02}.
\bibitem{higgshosotani1}
H. Hatanaka, K. Ohnishi,M. Sakamoto and K. Takenaga,
\PTPM{107}{1191}{02}, \PTPM{110}{791}{03}.
\bibitem{higgshosotani2}
K. Ohnishi and M. Sakamoto, \PLBM{486}{179}{00};
H. Hatanaka, S. Matsumoto, K. Ohnishi and M. Sakamoto,
\PRDM{63}{105003}{01}.
\bibitem{rs}
L. Randall and R. Sundrum, \PRL{83}{4690}{99}. 
\bibitem{manybrane}
H. Hatanaka, M. Sakamoto, M. Tachibana and K. Takenaga, \PTP{102}{1213}{99}.
\bibitem{nagasawasakamoto}
T. Nagasawa and M. Sakamoto, \PTPM{112}{629}{04}.
\bibitem{sakamototakenaga1}
M. Sakamoto and K. Takenaga, \PRDM{75}{045015}{07}.
\bibitem{KL}
D. A. Kirzhnits, A. Linde, \PLB{42}{471}{72}.
\bibitem{dolanjackiw}
L. Dolan and R. Jackiw, \PRD{9}{3320}{74}.
\bibitem{weinberg}
S. Weinberg, \PRD{9}{3357}{74}.
\bibitem{cosmology}
R. Mohapatra and G. Senjanovic, \PRL{42}{1651}{79}; 
\PRD{20}{3390}{79}; G. R. Dvali, A. Melfo and G. Senjanovic; 
\PRL{75}{4559}{95}; G.R. Dvali and 
K. Tamvakis, \PLB{378}{141}{96}; G. Bimonte 
and G. Lozano, \NPB{460}{155}{96}, M. B. Pinto and R. O. Ramos,
\PRDM{61}{125016}{00}; M. B. Pinto, R. O. Ramos and
J. E. Parreira, \PRDM{71}{123519}{05}.
\bibitem{gpy}
D. J. Gross, R. D. Pisarski and L. G. Yaffe, \RMP{53}{43}{81}.
\bibitem{sakamototakenaga2}
M. Sakamoto and K. Takenaga, \PRDM{76}{085016}{07}. 
\bibitem{weiss}
N. Weiss, \PRD{24}{475}{81}, \PRD{25}{2667}{82}.
\bibitem{KL2001}
C. P. Korthals and M. Laine, \PLBM{511}{269}{01}.
\bibitem{FP}
K. Farakos and P. Pasipoularides, \NPBM{705}{92}{05}.
\bibitem{marutake}
N. Maru and K. Takenaga, \PRDM{72}{046003}{05};
\PRDM{74}{015017}{06}.
\bibitem{pomarol}
A. Delgado, A. Pomarol and M. Quiros, \PRD{60}{095008}{99}.
\bibitem{takenaga}
K. Takenaga, \PLBM{570}{244}{03}.
\bibitem{hatakugo}
H. Hata and T. Kugo, \PRD{21}{3333}{80}.
\bibitem{ginsparg}
P. H. Ginsparg, \NPB{170}{388}{80}.
\bibitem{bn}
E. Braaten and A. Nieto, \PRD{51}{6990}{95}.
\bibitem{littleHiggs}
J. R. Espinosa, M. Losada and A. Riotto, \PRDM{72}{043520}{05}. 
\bibitem{models}
A. T. Davies and A. McLachlan, \NPB{317}{237}{89}, 
A. McLachlan, \NPB{338}{188}{90}, 
J. E. Hetrick and C. L. Ho, \PRD{40}{4085}{89}, C. L. Ho 
and Y. Hosotani, \NPB{345}{445}{90}, 
A. McLachlan, \NPB{338}{188}{90}, H. Hatanaka, \PTP{102}{407}{99}, 
K. Takenaga, \PLB{425}{114}{98}; \PRD{58}{026004}{98};
{\bf 66} 085009 (2002); \PLBM{570}{244}{03}; 
N. Haba, K. Takenaga and T. Yamashita, \PLBM{605}{355}{05}.
\bibitem{sakatakenew}
M. Sakamoto and K. Takenaga (work in progress).
\end{thebibliography}
\end{document}